\documentclass[twocolumn]{aastex631}
\submitjournal{ApJ Letter}
\received{September 18, 2024}
\revised{November 12, 2024}
\accepted{November 19, 2024}

\shorttitle{Fe K$\alpha$ orbital modulation of Cen X-3 with \textit{XRISM}/\textit{Resolve}}

\begin{document}
\title{
Detection of the orbital modulation of Fe K$\alpha$ fluorescence emission in Centaurus X-3
using the high-resolution spectrometer \textit{Resolve} onboard \textit{XRISM}
}
\correspondingauthor{Yuto Mochizuki}
\email{mochizuki@ac.jaxa.jp}

\author[0000-0003-3224-1821]{Yuto Mochizuki}
\affil{Institute of Space and Astronautical Science (ISAS), Japan Aerospace Exploration Agency (JAXA),\\3-1-1 Yoshinodai, Chuo-ku, Sagamihara, Kanagawa 252-5210, Japan}
\affil{Department of Astronomy, Graduate School of Science, The University of Tokyo, 7-3-1 Hongo, Bunkyo-ku, Tokyo 113-0033, Japan}

\author[0000-0002-9184-5556]{Masahiro Tsujimoto}
\affil{Institute of Space and Astronautical Science (ISAS), Japan Aerospace Exploration Agency (JAXA),\\3-1-1 Yoshinodai, Chuo-ku, Sagamihara, Kanagawa 252-5210, Japan}

\author[0009-0007-2283-3336]{Richard L. Kelley}
\affil{X-ray Astrophysics Laboratory, Code 662, NASA Goddard Space Flight Center, 8800 Greenbelt Rd., Greenbelt, MD 20771, USA}

\author[0000-0002-5488-1961]{Bert Vander Meulen}
\affil{Sterrenkundig Observatorium, Universiteit Gent, Krijgslaan 281 S9, 9000 Gent, Belgium}

\author[0000-0003-1244-3100]{Teruaki Enoto}
\affil{Department of Physics, Kyoto University, Kitashirakawa-Oiwake-cho, Sakyo-ku, Kyoto, 606-8502, Japan}

\author{Yutaro Nagai}
\affil{Department of Physics, Kyoto University, Kitashirakawa-Oiwake-cho, Sakyo-ku, Kyoto, 606-8502, Japan}

\author[0000-0002-1065-7239]{Chris Done}
\affil{Centre for Extragalactic Astronomy, Department of Physics, University of Durham, South Road, Durham DH1 3LE, UK}

\author[0000-0002-1131-3059]{Pragati Pradhan}
\affil{Department of Physics, Embry-Riddle Aeronautical University: Prescott, Arizona, USA}

\author[0000-0003-3057-1536]{Natalie Hell}
\affil{Lawrence Livermore National Laboratory: Livermore, CA, USA}

\author[0000-0002-4656-6881]{Katja Pottschmidt}
\affil{University of Maryland Baltimore County, 1000 Hilltop Circle, Baltimore, MD 21250, USA}
\affil{CRESST \& Astroparticle Physics Laboratory, Code 661, NASA Goddard Space Flight Center, 8800 Greenbelt Rd., Greenbelt, MD 20771, USA}

\author[0000-0002-5352-7178]{Ken Ebisawa}
\affil{Institute of Space and Astronautical Science (ISAS), Japan Aerospace Exploration Agency (JAXA),\\3-1-1 Yoshinodai, Chuo-ku, Sagamihara, Kanagawa 252-5210, Japan}
\affil{Department of Astronomy, Graduate School of Science, The University of Tokyo, 7-3-1 Hongo, Bunkyo-ku, Tokyo 113-0033, Japan}

\author[0000-0001-9735-4873]{Ehud Behar}
\affil{Physics Department, Technion, Haifa 32000, Israel}

\begin{abstract} 
 The Fe K$\alpha$ fluorescence line emission in X-ray spectra is a powerful diagnostic
 tool for various astrophysical objects to reveal the distribution of cold matter around
 photo-ionizing sources. The advent of the X-ray microcalorimeter onboard the
 \textit{XRISM} satellite will bring new constraints on the emission line. We present
 one of the first such results for the high-mass X-ray binary Centaurus X-3, which is
 composed of an O-type star and a neutron star (NS). We conducted a 155~ks observation
 covering an entire binary orbit. A weak Fe K$\alpha$ line was detected in all orbital
 phases at an equivalent width (EW) of 10--20~eV. We found for the first time that its
 radial velocity (RV) is sinusoidally modulated by the orbital phase. The RV amplitude
 is 248 $\pm$ 13~km~s$^{-1}$, which is significantly smaller than the value
 (391~km~s$^{-1}$) expected if the emission is from the NS surface, but is
 consistent if the emission takes place at the O star surface. We discuss several
 possibilities of the line production site, including the NS surface, O star surface, O
 star wind, and accretion stream from the O star to the NS. We ran radiative transfer
 calculation for some of them assuming spherically-symmetric density and velocity
 profiles and an isotropic distribution of X-ray emission from the NS. None of them
 explains the observed EW and velocity dispersion dependence on the orbital phase,
 suggesting that more elaborated modeling is needed. In other words, the present
 observational results have capability to constrain deviations from these assumptions.
\end{abstract}


\section{Introduction}\label{s1}
In X-ray spectroscopy, the Fe K$\alpha$ line at 6.4~keV is a powerful probe to
investigate the distribution of cold matter around X-ray sources. It is fluorescence
emission of neutral or low-ionized Fe following the inner-shell ionization by photons
above the binding energy of 7.12~keV. Due to the large cosmic abundance and high
fluorescence yield of iron, the Fe K$\alpha$ emission line stands out in X-ray spectra
of a wide range of objects, including active galactic nuclei, X-ray binaries,
cataclysmic variables, and normal stars. In many cases, however, the location of the
reprocessing cold matter is a matter of debate. Thus, it is important to characterize
the Fe K$\alpha$ emission line complex with observations of higher spectral resolution
and signal-to-noise ratio.

The high-energy transmission grating (HETG) spectrometer onboard the \textit{Chandra}
X-ray Observatory \citep{canizares2005} was the first to have the capability to resolve
the line profile at an energy resolution $R \equiv E/\Delta E \sim 167$ and an effective
area $A_{\mathrm{eff}} \sim 20$~cm$^{2}$ at
6.4~keV\footnote{\url{https://cxc.cfa.harvard.edu/proposer/POG/html/index.html}},
providing information besides its equivalent width (EW) available with spectrometers
with lower energy resolution. Two systematic studies were made on the narrow Fe
K$\alpha$ emission from bright X-ray binaries using HETG
\citep{torrejon2010,tzanavaris2018}, in which the former argued that the line is
unresolved while the latter argued that it has a FWHM of
$\sim$5000~km~s$^{-1}$. These loose constraints are the best that could be achieved even
for bright X-ray sources before \textit{XRISM}.

This has changed drastically with the advent of \textit{XRISM} X-ray
microcalorimeter. With an improved $R \sim 1400$ and $A_{\mathrm{eff}} \sim
174$~cm$^{2}$ at 6.4~keV, the \textit{Resolve} instrument \citep{ishisaki2022} onboard
the \textit{XRISM} satellite \citep{Tashiro2020} will reveal new aspects of the
emission such as tighter constraints of the line profile, its changes, and the timing
delays of line photons compared to continuum photons. These abilities are clearly
demonstrated in X-ray binaries hosting a neutron star (NS), by determining how the
emission is modulated both by the orbital and spin rotations.

\medskip

Here, we present the first such result for Centaurus X-3 (Cen X-3), which is an
eclipsing high-mass X-ray binary (HMXB) consisting of a NS and an O6.5 II-III star. The
NS was discovered in X-rays \citep{chodil1967} followed by the O star companion in the
optical \citep{krzeminski1974}. The NS is the first X-ray pulsar discovered as such with
a spin period $P_{\mathrm{spin}}=4.8$~s \citep{Giacconi1971}. The system parameters are
well known. Those we use in this paper are: the orbital period
$P_{\mathrm{orb}}=2.08$~days and its change $dP_{\mathrm{orb}}/P_{\mathrm{orb}}=-1.8
\times 10^{-6}$~yr$^{-1}$ from the X-ray eclipse timing \citep{schreier1972,klawin2023},
the mass of the NS and O star $M_{\mathrm{NS}}=1.21 \pm 0.21 M_{\odot}$ and
$M_{\mathrm{O}}=20.5 \pm 0.7 M_{\odot}$ from the optical radial velocity (RV)
\citep{ash1999,VanderMeer2007} and X-ray pulse timing \citep{Kelley1983} measurements
modulated by the orbital motion, the semi-major axis length $a=$42.1~lt-s
\citep{bildsten1997}, the inclination angle $i=70.2 \pm 2.7$~degree \citep{ash1999}, the
elipticity $e = 0.00$ \citep{Kelley1983}, and the distance of $D=$~6.4~kpc from a
parallax measurement \citep{arnason2021}.

\begin{figure*}[!hbtp]
 \begin{center}
 \includegraphics[width=1.0\textwidth,clip]{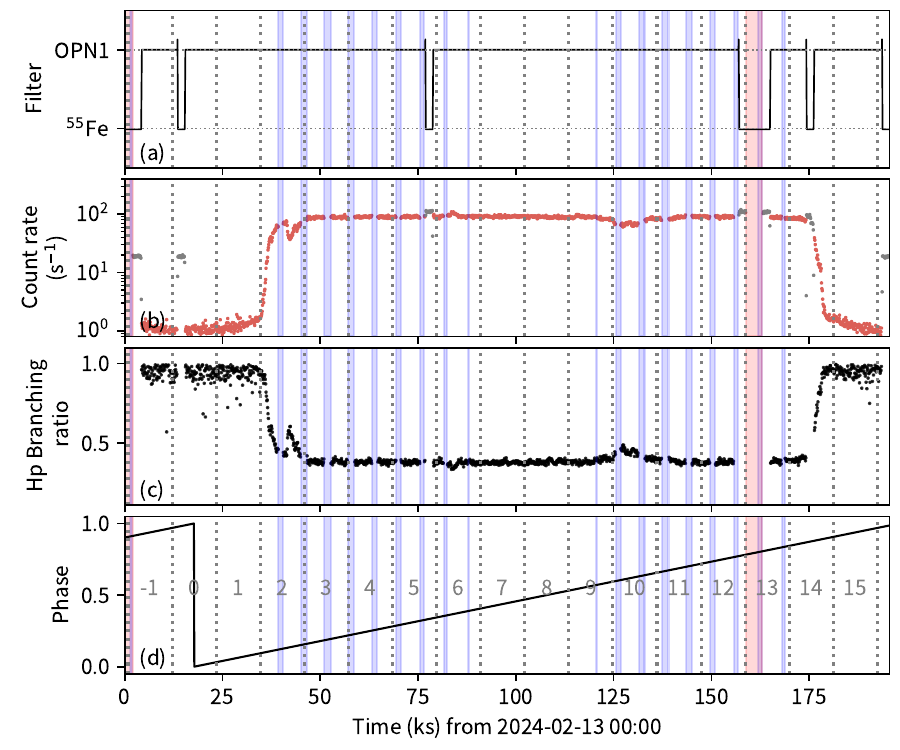}
 \caption{(a) FW positions, (b) 2--12~keV count rate binned at 100~s, (c) the fraction
  of Hp events, and (d) orbital phase \citep{falanga2015}. The observation was
  interrupted by South Atlantic Anomaly passages (blue) and ADR recycles (red). The FW
  $^{55}$Fe was illuminated several times as shown in (a) during on-source integration
  time, leading to an additional count rate of 20~s$^{-1}$.}
  \label{f01}
\end{center}
\end{figure*}

\section{Observation and Data Reduction}\label{s2}
\subsection{Instrument}
\textit{Resolve} hosts an array of 6$\times$6 X-ray microcalorimeters based on the HgTe
absorber and the Si thermister thermally anchored to the 50~mK stage at a thermal time
constant of $\sim$3.5~ms \citep{kilbourne18b}, which is maintained with two-stage
adiabatic demagnetization refrigerators (ADRs; \citealt{shirron18}) from the 1.2~K stage
by the depressurized superliquid helium stored in the cryostat \citep{ezoe2020}. One of
the pixels (pixel 12) is displaced from the array for calibration purposes. Individual
pixels have their own energy gain variation in orbit, which is tracked and corrected
\citep{porter16} using the $^{55}$Fe sources placed in one of the filter wheel (FW)
windows \citep{devries18}.

The gate valve (GV) has not yet been opened in orbit, despite several initial
attempts. It is an apparatus that keeps the cryostat leak-tight prior to launch and is
intended to be opened in orbit. The X-ray transmissive window of the GV is made with Be
of a $\sim$250~$\mu$m thickness \citep{midooka2021a}, which restricts the bandpass to
above $\sim$2~keV. However, this is fortuitous for observations of bright sources like
the one presented here, since the photon count rates are attenuated in the appropriate
dynamic range. Other details of the instrument can be found in \citet{sato2023}.

The X-ray microcalorimeter spectrometer excels over CCD-based spectrometers besides $R$
and $A_{\mathrm{eff}}$. Unlike the grating spectrometers onboard the \textit{Chandra}
and \textit{XMM-Newton} observatories, the calibration sources are intermittently
illuminated during observations to achieve an energy determination accuracy of $\lesssim
1$~eV. The HgTe absorber provides bandpass to 12~keV and beyond, which is inaccessible
with Si absorbers. The event time tagging is accurate to $\sim$5~$\mu$s
\citep{omama2022} for the need to resolve event pulse shape in time in orbit, which
is better than conventional X-ray CCD spectrometers by many orders. The anti-coincidence
detector discriminates particle-induced events efficiently to reduce the background rate
to only one event per spectral resolution ($\sim$5~eV) per 100~ks
\citep{kilbourne18}. All of these contribute to the improved characterization of the Fe
K$\alpha$ line and its underlying continuum photons, as will be presented below.

\subsection{Observation}
The observation (sequence number 300003010) was made from 2024 February 12 23:56:04 to
February 15 06:19:04 for a telescope time of 196~ks as a performance verification
program. The observation efficiency is very high of 79\% for a satellite in a low Earth
orbit, as we chose a season when the target is visible without Earth occultation. The
\textit{XRISM} observation covered an entire binary cycle starting from one eclipse to
the next (Fig.~\ref{f01}d). In this paper, we focus on the \textit{Resolve} data taken outside the eclipses.

Figure~\ref{f01} (b) shows the light curve in the 2--12~keV band. The eclipses are seen
at the beginning and end of the observation. The count rate outside the eclipse shows an
average rate of 54--112~s$^{-1}$ binned at 100~s. At these rates, the background is
negligible \citep{mochizuki2024} and the event loss due to the limited computing
resources in orbit does not occur \citep{mizumoto2022}.

The X-ray light curve is very stable, which is often seen when the X-ray luminosity is
high of $\sim 10^{37}$~erg~s$^{-1}$ as in the 2006 June observation with
\textit{XMM-Newton} \citep{sanjurjo-ferrin2021} and the 2022 July observation with
\textit{IXPE} \citep{weisskopf2022}. In other observations, the X-ray light curve is
much more variable \citep{suchy2008,naik2011,tamba2023}.

\subsection{Data reduction}
We retrieved the pipeline products processed with a script version 03.00.011.008. We
reduced the data using the \texttt{HEASoft} package build 7 released for the
\textit{XRISM} science team. For timing analysis, we used all events screened in the
standard processing, which amounts to 9.3 million events. For spectroscopy analysis, we
applied further event screening to remove events below 0.3~keV, those with too fast or
slow pulse rise times for their energy \citep{mochizuki2024}, those with grades other
than Hp, and those recorded in pixel 27, which is reported to have an irregular gain
trend. A total of 3.6 million events were left. Here, the Hp grade is given for events
without any other events within 70.72~ms in the same pixel \citep{ishisaki2018}. They
have the highest accuracy in energy determination and are well calibrated for
spectroscopy as of writing. The Hp branching ratio depends on the concurrent count rate
\citep{seta12}. The average ratio out of the eclipse is $\sim$0.39 (Fig.~\ref{f01}c).

The detector and telescope response files were generated using the \texttt{rslmkrmf} and
\texttt{xaarfgen} tools. No background spectrum was subtracted as it is
negligible. Events were divided in time into 16 pieces per orbit that covered
$\phi_{\mathrm{orb}} = [\frac{i_{\mathrm{orb}}-0.5}{16},
\frac{i_{\mathrm{orb}}+0.5}{16}]$, where $i_{\mathrm{orb}} \in\{-1,0,1,...,15\}$
(Fig.~\ref{f01}d). Among them, $i_{\mathrm{orb}} \in\{3,4,5,...,13\}$ are out of the
eclipse. A large fraction of the exposure time was lost for $i_{\mathrm{orb}}=13$ due to
ADR recycling. The quoted errors hereafter represent 1 $\sigma$ statistical uncertainty.

\section{Analysis}\label{s3}
\subsection{Spectra}\label{s3-1}
\begin{figure*}[!hbtp]
 \centering
 \includegraphics[width=1.0\textwidth,clip]{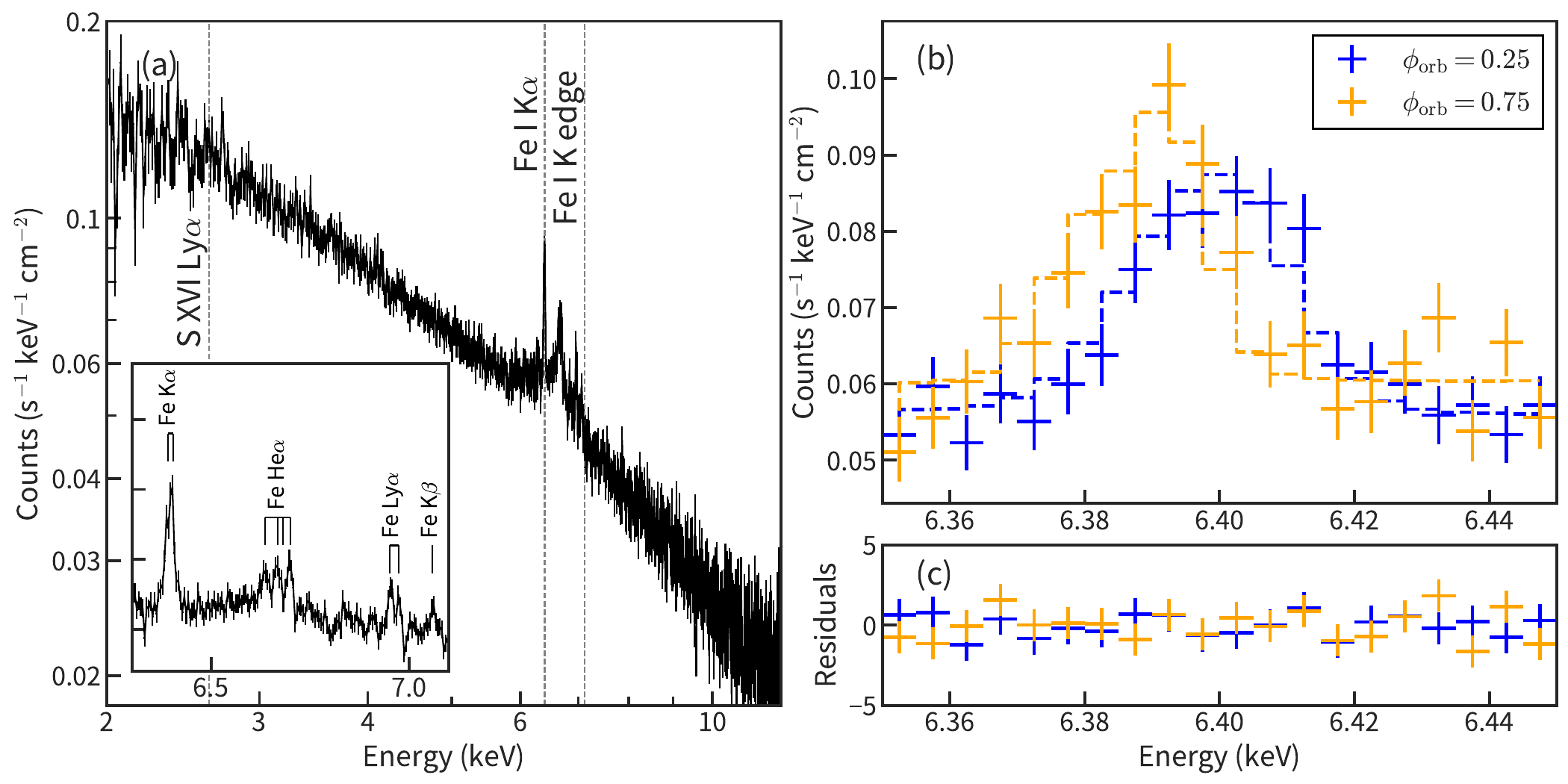}
 \caption{(a) \textit{Resolve} spectra out of the eclipse in the 2-12~keV band. A
 close-up view of the Fe K band is given in the inset. (b) Fe K$\alpha$ spectra and
 their best-fit models and (c) residuals to the fit at $\phi_{\mathrm{orb}}=$0.25 (blue)
 and 0.75 (orange). Compare to Figure 4 (a) in \citet{wojdowski2003a} for previously the
 best spectra with the \textit{Chandra} grating spectrometer.}
 \label{f02}
\end{figure*}

Figure~\ref{f02} (a) shows the broadband spectrum with \textit{Resolve} integrated over
the entire off-eclipse phase. Upon the hard continuum emission, emission lines of highly
ionized S and Fe as well as the fluorescence of neutral or low-ionized Fe are
evident. Fe K band is particularly rich; Fe XXV He$\alpha$ and Fe XXVI Ly$\alpha$ lines
are resolved. Their resonance lines exhibit a clear signature of the P Cygni profile.

\begin{figure*}[!hbtp]
 \centering \includegraphics[width=1.0\textwidth,clip]{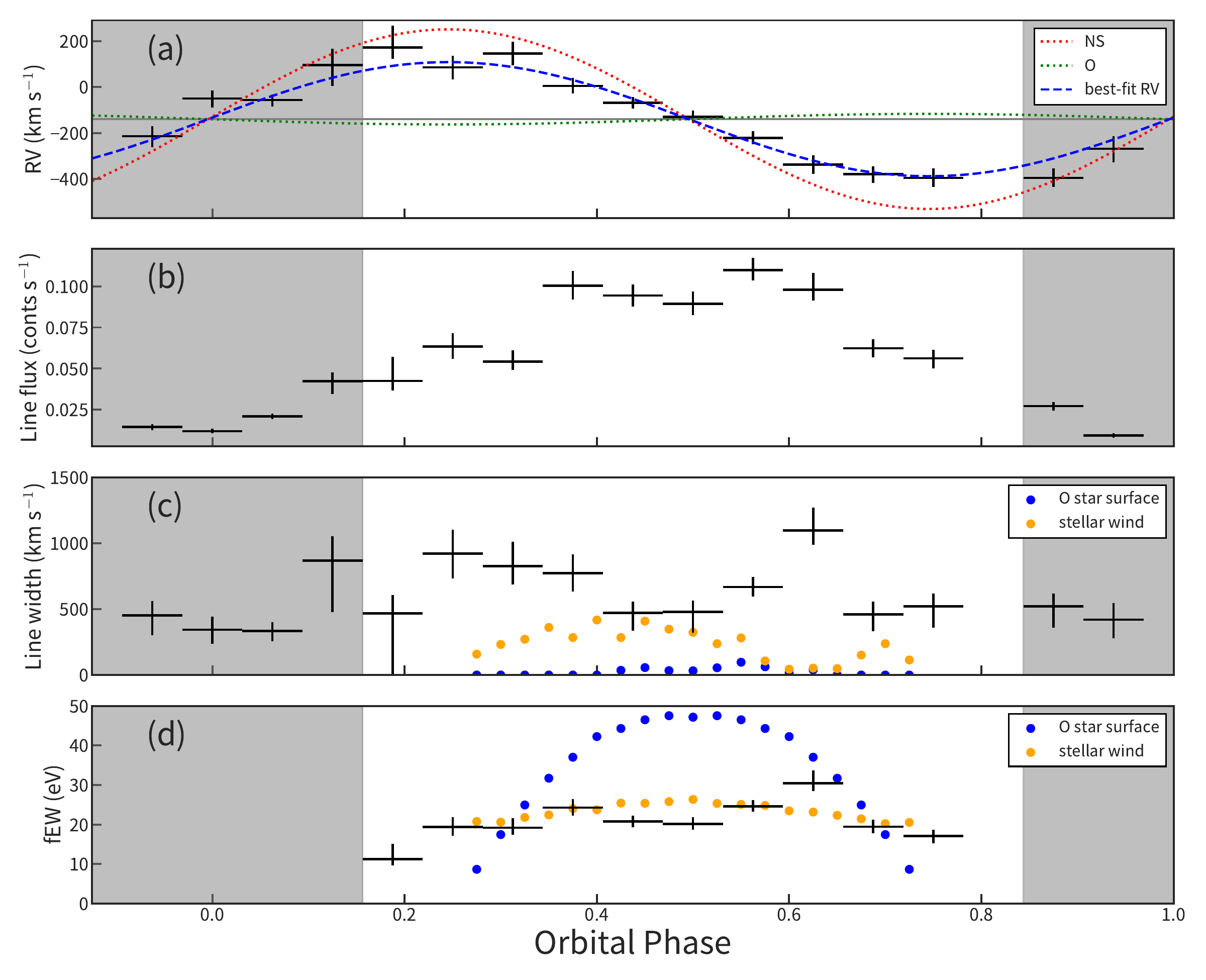}
 \caption{Best-fit values of (a) Fe K$\alpha$ line center, (b) flux, and (c) width as
 well as (d) fluorescence EW as a function of $\phi_{\mathrm{orb}}$. Shaded areas
 indicate eclipses. In (a), the data are compared to the RV curve of the NS (red), O
 star (green), and the best fit to the data outside the eclipse (blue). Positive RVs are
 in the approaching direction. Results of the radiative transfer calculation are
 compared in (c) and (d) for the O star surface (blue) and stellar wind (orange)
 as the site of Fe K$\alpha$ production.}
 \label{f04}
\end{figure*}

Figure~\ref{f02} (b) shows the narrow-band spectrum of the Fe K$\alpha$ line at two
opposite orbital phases. The center energy is significantly displaced. We constructed
the narrow-band spectrum at each $\phi_{\mathrm{orb}}$ bin and fitted the Fe K$\alpha$
line with an empirical model consisting of a handful of Lorentzian profiles
\citep{holzer1997}. The free parameters are the center energy, width, and normalization
for the set of the lines, and the index and normalization of the underlying power-law
continuum. Fe K$\alpha$ lines broader than the one presented here may additionally
exist, but it is not required in the narrow-band fitting. The best-fit values are shown
in Figure~\ref{f04} (a)--(c).

To validate the result, we also fitted the Mn K$\alpha$ and K$\beta$ lines of the
electron capture decay of $^{55}$Fe during the FW $^{55}$Fe illuminations in the same
data set. The \citet{holzer1997} model with corrections based on the microcalorimeter
data\footnote{\url{https://heasarc.gsfc.nasa.gov/docs/hitomi/calib/hitomi_caldb_docs.html}}was
used. The energy of the Mn K$\alpha$ (Mn K$\beta$) line is accurate to 0.06 $\pm$ 0.01
(0.31 $\pm$ 0.04~eV) equivalent to 3.0 $\pm 0.6$ (15 $\pm$ 2) km~s$^{-1}$.  The
broadening besides the instrumental line spread function is $<5\times10^{-4}$
($<10^{-4}$)~eV equivalent to $<2\times10^{-2}$ ($<10^{-2}$)~km~s$^{-1}$.

The RV dependence on $\phi_{\mathrm{orb}}$ (Fig.~\ref{f04}a) shows a clear sinusoidal
pattern. We fitted the data outside of the eclipse with a sine curve and obtained the
best-fit RV amplitude of 248$\pm$13~km~s$^{-1}$, the RV offset of
--140$\pm$4~km~s$^{-1}$, and the phase offset of $0.004 \pm 0.001$. The phase offset is
consistent with null. The RV offset is larger than the arithmetic sum of the gain
uncertainty and \textit{XRISM} motion around the Solar barycenter
($+$15~km~s$^{-1}$). We attribute the rest to the proper motion of Cen X-3 (--39
km~s$^{-1}$; \citealt{hutchings1979}) and other factors.

The line flux (Fig.~\ref{f04}b) also exhibits a change as a function of
$\phi_{\mathrm{orb}}$ with approximately a single peak profile outside of the
eclipse. During the eclipse, the line flux is reduced but not null, which is
consistent with the previous measurements \citep{aftab2019}.
The line width (Fig.~\ref{f04}c) is significantly broader than null across
$\phi_{\mathrm{orb}}$ with 10--20~eV corresponding to 500--1000~km~s$^{-1}$ in FWHM. If
the broadening did not exist, the K$\alpha^{1}$ and K$\alpha^{2}$ lines would have been
easily resolved (Fig.~\ref{f02}b).

We evaluated the equivalent width (EW) as well as the fluorescence equivalent width
(fEW) of the Fe K$\alpha$ line (Fig.~\ref{f04}d). Here, fEW is defined as the line
intensity divided by the continuum flux at 7.10--7.11~keV just below the Fe K edge
energy. The metric is more physical for the fluorescence line than the conventional EW
that uses the continuum flux at the line energy, but fEW and EW are different only by a
small factor ($\sim$1.1). The fEW can be assessed for the first time with
\textit{Resolve} without being contaminated by the associated Fe K$\beta$ line at
7.05~keV. The fEW shows a relatively flat distribution over $\phi_{\mathrm{orb}}$ out of
the eclipse and a much larger value during the eclipse.

\subsection{Timing}\label{s3-2}
\begin{figure*}[!hbtp]
 \centering
 \includegraphics[width=\textwidth,clip]{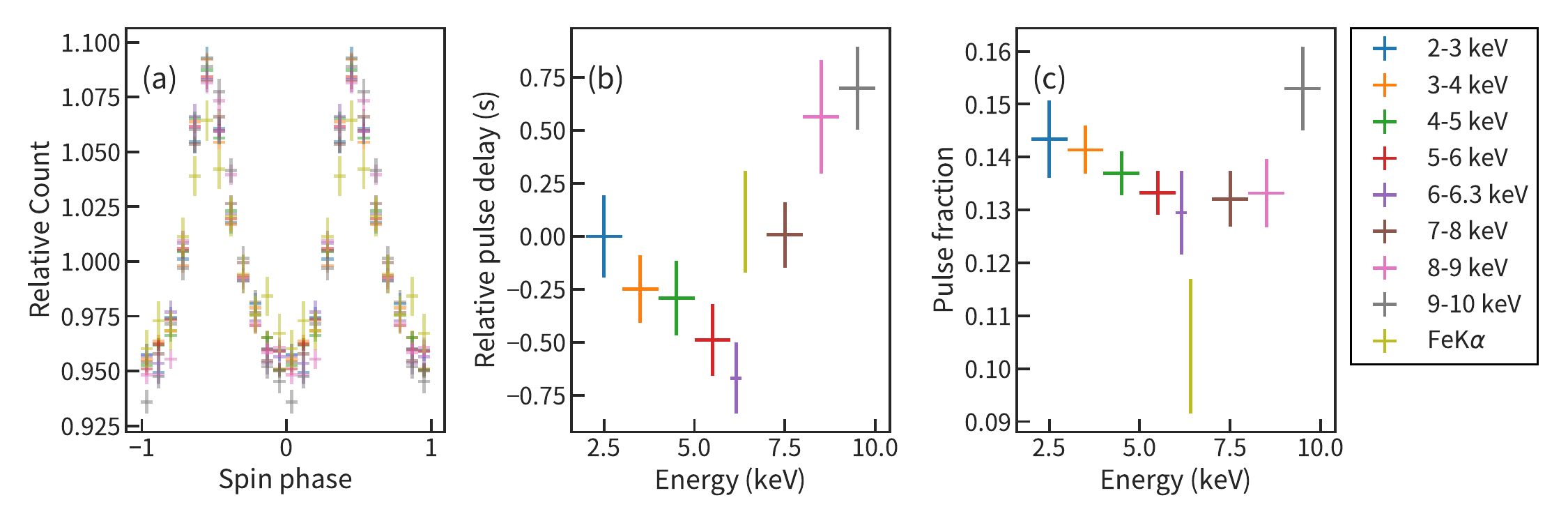}
 \caption{(a) Pulse shape folded by $P_{\mathrm{spin}}$, (b) phase delay relative to the
 lowest energy band as a function of energy, and (c) pulse fraction as a function of
 energy. The Fe K$\alpha$ count rate is from 6.35--6.45 keV.}
 \label{f06}
\end{figure*}

For all events, we corrected for the arrival times due to the \textit{XRISM} motion
around the barycenter of the Solar system and the neutron star motion around the center
of mass (CoM) in the Cen X-3 system. We then derived the spin period
$P_{\mathrm{spin}}=4.79784$~s using the epoch folding search. The pulse profile is made
for different broad energy bands as well as a narrow band containing the Fe K$\alpha$
(6.35--6.45 keV) and is normalized to the average (Fig.~\ref{f06}a). Here, the use of
all event grade is important, as the grade branching ratio changes as a function of pulse
phase.

The pulse profile is similar, but different, among the different energy bands. We derived
the pulse delay relative to the lowest energy band by cross-correlating the profiles
(Fig.~\ref{f06}b). We also derived the pulse fraction as the peak-to-peak value of the
normalized profile (Fig.~\ref{f06}c). By comparing between the 7--8~keV (ionizing
photons) and Fe K$\alpha$ band (fluorescent photons), we found no significant phase
delay but a significant drop in pulse fraction in Fe K$\alpha$. The decrement of the
pulse fraction at the Fe K band was reported in low-resolution timing spectroscopy
\citep{ferrigno2023}, which is now revealed to be associated to the Fe K$\alpha$
fluorescence line.

\section{Discussion}\label{s4}
In Cen X-3, the Roche lobe is considered to be filled, and the accretion disk is formed
around the NS. The matter from the O star accretes through the first Lagrangian point
(L1) via the accretion stream to the outer part of the accretion disk, which is truncated at
an inner radius, where magnetic pressure dominates. See also Figure~10 in
\citet{tsygankov2022}.

\begin{figure}[!hbtp]
 \centering
 \includegraphics[width=1.0\columnwidth,clip]{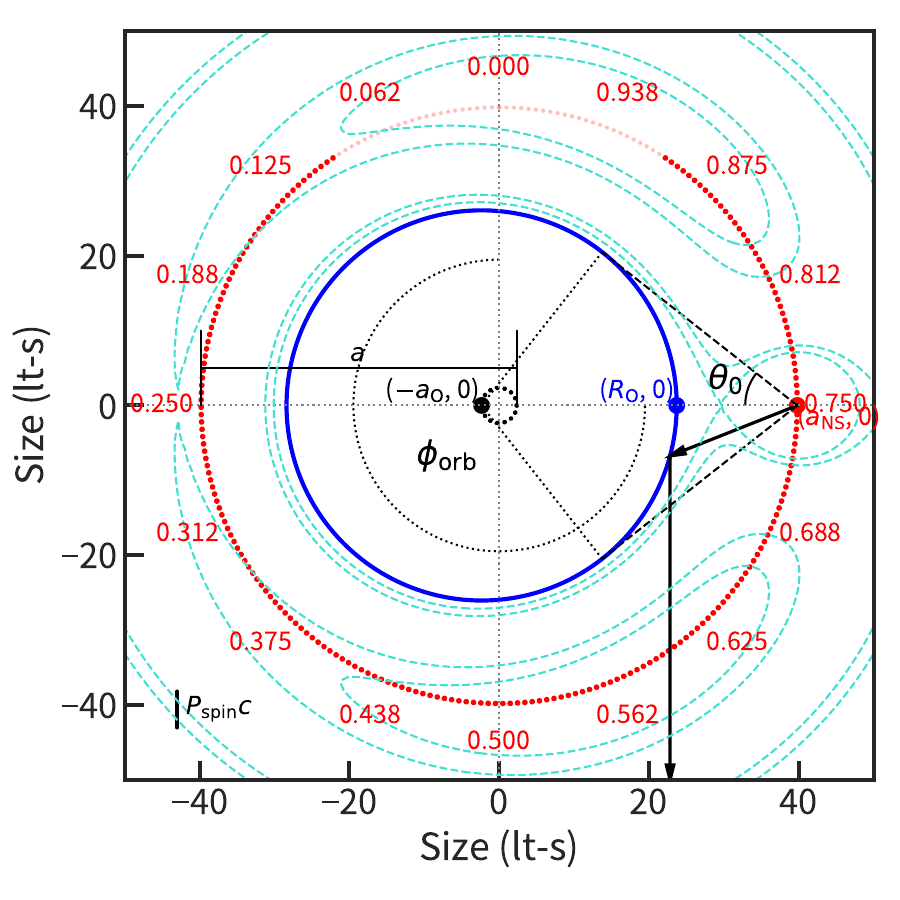}
 \caption{Schematic view and the radiative transfer calculation setup at
 $\phi_{\mathrm{orb}}=0.75$. The CoM is placed at the origin and the orbit is projected
 on its plane. The line of sight is toward $+y$-axis direction. The locations of the NS
 (red), O star (black), and an O star surface closest to the NS (blue) are shown with
 filled circles of their colors. The NS and O star orbits are shown with dotted
 circles. For the NS orbit, the eclipsed part is shown in transparency and
 $\phi_{\mathrm{orb}}$ is shown in text. The Roche potential is shown with the turquoise
 contours. The scale length of $P_{\mathrm{spin}}c$ is shown at the bottom left. For the
 modeling, the O star surface is approximated by a sphere with a radius of
 $R_{\mathrm{O}}$ in blue, which subtends an angle $\theta_0$ seen from the NS. A
 reflection path is shown with black arrows.}
 \label{f07}
\end{figure}

The system scale is shown in Figure~\ref{f07}. The radius of the O star is approximated by the
effective Roche-lobe radius $R_{\mathrm{O}} = 0.62 a$ for the mass ratio
$M_{\mathrm{NS}}/M_{\mathrm{O}}=0.059$ \citep{eggleton83}. This is consistent with the
value ($0.62a$) derived from the eclipse duration \citep{mouchet1980}. When viewed from
the NS, the O star subtends $\Delta \Omega_{\mathrm{O}} = 2\pi (1- \cos{\theta_0})$ str,
where $\theta_0 = \arcsin{(R_{\mathrm{O}}/a)}$. For $R_{\mathrm{O}} = 0.62a$, $\Delta
\Omega_{\mathrm{O}} / 4\pi= 0.1$. The light travel distance for the spin period is
$P_{\mathrm{spin}}c = 0.11a$. Below, we consider several possibilities for the
production site of the Fe K$\alpha$ line.

First, it has been postulated that the most plausible location of the reprocessing cold
matter is the vicinity of NS for two major reasons: (1) the line intensity drops during
the eclipse \citep{nagase1992,Ebisawa1996,naik2011} using the data taken with
\textit{Ginga}, \textit{ASCA}, and \textit{XMM-Newton} and (2) the line photons are
delayed in the pulse phase only by 0.39~ms \citep{kohmura2001} corresponding to $\sim
10^{-5} a$. If the reprocessing matter is so close to the NS, we would expect the RV of
the line to be modulated by the NS motion with an amplitude of $2\pi a_{\mathrm{NS}}
\sin{i} /P_{\mathrm{orb}} =$391 $\pm$8~km~s$^{-1}$, where $a_{\mathrm{NS}}=a
M_{\mathrm{O}}/(M_{\mathrm{NS}}+M_{\mathrm{O}})$ is the radius of the NS orbit around
the CoM of the binary (red circle in Fig.~\ref{f07}). The actual measurement is
significantly smaller (Fig.~\ref{f04}). Also, it does not explain the decrement of the
pulse fraction at the Fe K$\alpha$ band (Fig.~\ref{f06}c).

Second, we consider the surface of the O star as the site of reprocessing. This would
naturally explain the RV amplitude if the O star surface rotates synchronously with the
NS orbital motion. To take into account three-dimensional effects, we performed a
radiative transfer calculation using \texttt{SKIRT} \citep{vandermeulen2023}, which is a
three-dimensional Monte Carlo solver. The setup is shown in Figure~\ref{f07}. We assumed
that the O star fills the Roche lobe and the reflection takes place on its skin surface
with a density of $10^{18}$~cm$^{-3}$ but with no velocity. We launched photons
isotropically from the point-like NS placed at varying $\phi_{\mathrm{orb}}$. The
incident photons are sampled to follow a power-law spectral shape with a photon index of
1.5 in the 2--50~keV range. We considered the X-ray photon interactions with matter in
the O star for photoionization, fluorescence, and electron scattering. We derived the
fEW as a function of $\phi_{\mathrm{orb}}$ and compared it with the data
(Fig.~\ref{f04} d). A large discrepancy is found both in the absolute value and the
$\phi_{\mathrm{orb}}$ dependence. The contribution by the O star surface reflection is
unavoidable for its large subtended angle and should be dominant as long as the
radiation from the NS is isotropic. The result infers that the emission from the NS is
highly non-isotropic and most of its emission is not subtended by the O star
surface. The same result was found in another eclipsing HMXB Vela X-1 by
\citet{rahin2023}, who argued against the reprocessing on the companion star surface.

\begin{figure}[!hbtp]
 \centering
 \includegraphics[width=1.0\columnwidth,clip]{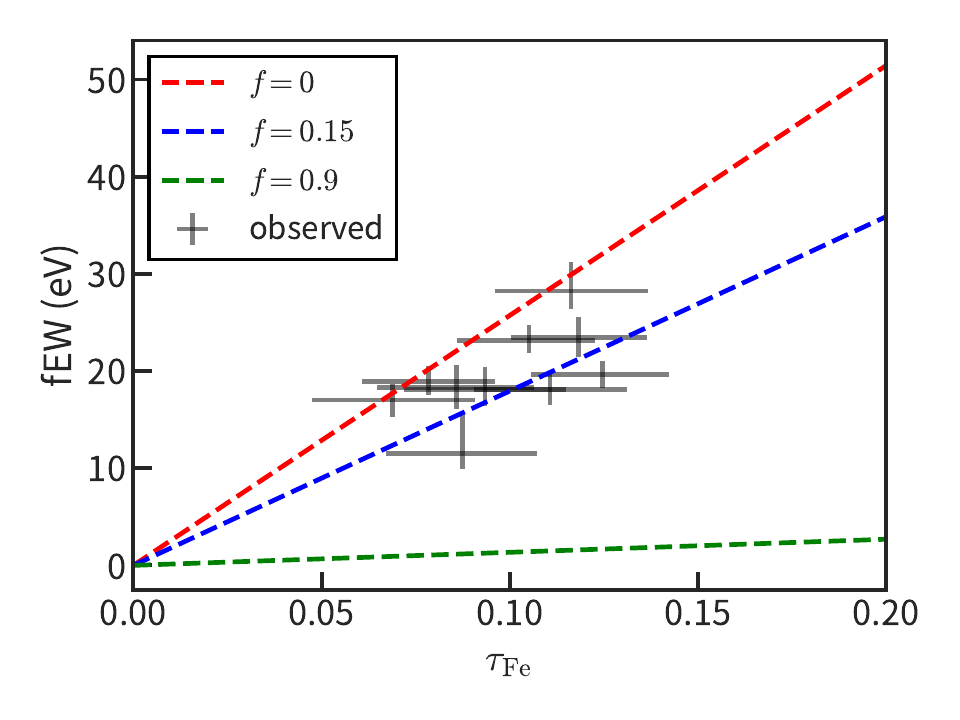}
 \caption{Relation between $\tau_{\mathrm{Fe}}$ and fEW. The observed results at
 different $\phi_{\mathrm{orb}}$ are shown with pluses, while the simulated results are
 shown with dotted lines of different colors for different $f \in \{0, 0.15, 0.9\}$.}
 \label{f18}
\end{figure}

Third, we consider cold matter in the line of sight. Its column density can be derived
from the Fe K edge depth ($\tau_{\mathrm{Fe}}$) at the edge energy
$E_{\mathrm{e}}=$~7.12~keV. The fEW should be proportional to $\tau_{\mathrm{Fe}}$ for
an optically-thin case. We used \texttt{SKIRT} in a different configuration, in which a
point-like NS placed at the center is surrounded by a shell with a thickness
corresponding to $\tau_{\mathrm{Fe}}$. Clumping of matter in the shell is also
considered with a parameter $f$, which is the fraction of total mass in
clumps. The remaining $1-f$ of mass is distributed uniformly in the shell.
Figure~\ref{f18} shows the observed and simulated results. The absolute value of fEW as
a function of $\tau_{\mathrm{Fe}}$ was not well reproduced with $f=0$ (uniform density
without clumps) but was with $f = 0.15$. This idea also explains some previous
observations \citep{naik2011,sanjurjo-ferrin2021}, in which Fe K$\alpha$ was found to be
much more intense with EW$>$100~eV. The variation of its intensity is correlated well
with the line-of-sight partial-covering column density derived from the broadband
spectral fitting. Note that the \textit{XRISM} observation was made when the
line-of-sight column is smaller than these observations.

\medskip

What is then the cold matter that produces the Fe K$\alpha$ line in the line of sight?
The present results require the following conditions. First, it has an RV amplitude
smaller than that of the NS (Fig.~\ref{f04}a). Second, it has a velocity dispersion of
500--1000~km~s$^{-1}$ in FWHM (Fig.~\ref{f04}c). Third, its density ($n$) should be high
enough so that the matter remains neutral or low ionized under a radiation field of
$L_{\mathrm{X}} \sim 10^{37}$~erg~s$^{-1}$; i.e., the ionization parameter ($\xi =
\frac{L_{\mathrm{X}}}{nr^2}$; \citealt{tarter1969}) should be less than $\sim$10
\citep{kallman2004}. This requires $n \gtrsim 10^{12} (r/a)^{-2}$~cm$^{-3}$, where $r$
is the distance from the NS to the cold matter. This further constrains the physical
thickness of matter ($d$) to be $d < 0.04a (r/a)^{2}$~cm from $\tau_{\mathrm{Fe}} = d n
A_{\mathrm{Fe}}\sigma_{\mathrm{K}}$, where $A_{\mathrm{Fe}}$ is the abundance of Fe
relative to H \citep{anders1989}, and $\sigma_{\mathrm{K}}$ is the photoelectric
absorption cross section at the edge of Fe K\footnote{We used the NIST standard
reference database 8.}. Fifth, $d$ needs to have some scatter of $\approx
P_{\mathrm{spin}}c$, so that the coherent pulse signal of the NS will spread over the
spin phase so that the reprocessed emission is consistent with the lack of phase delay
and the reduced pulse fraction (Fig.~\ref{f06}).

We list two possibilities. One is the stellar wind as argued for QV Nor
\citep{torrejon2015}. We ran another \texttt{SKIRT} simulation for a stellar wind having
a spherically-symmetric density and velocity distributions. We assumed a constant wind
mass loss rate $0.2 \times 10^{-6} M_{\odot}$~yr$^{-1}$, the radial velocity following
the $\beta=1.0$ law with the terminal velocity 1000 km s$^{-1}$, and the azimuthal
velocity with the same angular velocity with NS. This yielded a good match for fEW
dependence on $\phi_{\mathrm{orb}}$ (Fig.~\ref{f04} d), as the stellar wind is
distributed more largely than the O star surface. Yet, it fails to explain the observed
velocity dispersion (Fig.~\ref{f04} c).

The other is the accreting matter from the first Lagrangian point to the NS
illuminated by the non-isotropic incident emission from the NS. An accretion stream
trailing behind the NS is formed \citep{manousakis2015a,elmellah2018}, which is a
candidate for such a matter. This is in agreement with the recent measurement of the
polarization of the X-ray continuum emission, which is an independent and complementary
method to assess the reprocessing of the incident emission from the NS. Using
\textit{IXPE} \citep{weisskopf2022}, \citet{tsygankov2022} revealed polarized X-rays
with the degree of polarization modulated by the NS spin phase. They argued that the
polarization is made intrinsically as well as by reprocessing by the accreting matter to
the NS. 

\section{Conclusion}\label{s5}
We presented the result of X-ray microcalorimeter observations with the \textit{Resolve}
instrument onboard \textit{XRISM} using the data obtained during the performance
verification phase. Cen X-3 was observed for its entire orbital phase using a 155~ks
exposure at a high observing efficiency of 79\%. We focused on the Fe K$\alpha$ emission
line and revealed its new spectroscopic features as a function of the orbital phase and
new timing features as a function of the spin phase using the unique capability of the
\textit{Resolve} instrument. 

We examined several possibilities for the production site of the Fe K$\alpha$
emission, including the NS surface, O star surface, O star wind, and accretion stream
through L1. We ran radiative transfer calculation for some of them assuming spherically
symmetric distributions of the density and velocity as well as the isotropic
distribution of the X-ray illumination from NS. These assumptions are likely too
simplistic to explain the rich observed phenomenology, which point toward the need for
better modeling of such distributions. In other words, we can constrain them through the
present observational results.

We demonstrated, in a practical use case, that the X-ray microcalorimeter can reveal new
aspects of the Fe K$\alpha$ line, which is a powerful probe in many objects. As shown,
we only used a small fraction of the 9 million events. Other aspects of the data and
implications for their results will follow.

\begin{acknowledgments}
 The authors thank all those who contributed to the \textit{XRISM} mission. Richard
 Mushotzky provided useful comments on the manuscript. The anonymous reviewer
 pointed out that the stellar wind can be a production site of the Fe K$\alpha$
 emission. This work was supported by the JSPS Core-to-Core Program (grant number:
 JPJSCCA20220002) and by NASA under the award number 80GSFC21M0002. This research made
 use of the JAXA's high-performance computing system JSS3. Y\,M is financially supported
 by the JST SPRING program (grant number: JPMJSP2108), B\,V by the Fund for
 Scientific Research Flanders (FWO-Vlaanderen, project 11H2121N). Part of this work was
 performed under the auspices of the U.S. Department of Energy by Lawrence Livermore
 National Laboratory under Contract DE-AC52-07NA27344. The material is based upon work
 supported by NASA under award number 80GSFC21M0002. C\,D acknowledges support from STFC
 through grant ST/T000244/1.
\end{acknowledgments}

\vspace{5mm}
\facility{
\textit{XRISM} \citep{Tashiro2020} 
(
\textit{Resolve} \citep{ishisaki2022}
)
}
\software{
\texttt{Xspec (v.12.14.0h)} \citep{arnaud1996},
\texttt{SKIRT (v.9)} \citep{vandermeulen2023},
}
\bibliography{main}{}
\bibliographystyle{aasjournal}

\end{document}